\documentclass[11pt,a4paper]{article}
\usepackage{amsmath,amssymb,a4wide,graphicx}
\usepackage[numbers,sort&compress]{natbib}
\newcommand{\req}[1]{(\ref{#1})}
\newcommand{\vek}[1]{\boldsymbol{#1}}
\begin{document}
\title{Emergence of product differentiation from consumer
heterogeneity and asymmetric information}
\author{Linyuan L\"u$^{1,2}$, Mat\'u\v s Medo$^3$, Yi-Cheng
Zhang$^{2,3}$, Damien Challet$^{3,4}$}
\date{}
\maketitle

\begin{quote}
\small
$^1$ Department of Systems Science, School of Management,
Beijing Normal University, 100875~Beijing, China\\
$^2$ Lab of Information Economy and Internet Research,
University of Electronic Science and Technology of China,
610054~Chengdu, China\\
$^3$ Physics Department, University of Fribourg,
P\'erolles, 1700~Fribourg, Switzerland\\
$^4$ Institute for Scientific Interchange, via S. Severo 65,
10113~Turin, Italy
\end{quote}

\begin{abstract}
We introduce a fully probabilistic framework of consumer product
choice based on quality assessment. It allows us to capture many
aspects of marketing such as partial information asymmetry,
quality differentiation, and product placement in a~supermarket.
\end{abstract}

\section{Introduction}
The interests of vendors and customers seem antagonistic
\emph{a priori}, the former aiming at decreasing quality and
increasing price, whereas the latter wishing exactly the
opposite. The situation is fortunately more complex, the
interests of both sides being sometimes compatible. Intuitively,
a vendor may sell more items by increasing their perceivable
quality, making everybody happier. But the situation is more
subtle because of asymmetric information: the vendor knows much
better than his prospective customers the real quality of his
products. In Akerlof's famous Lemon Problem, the customers have
no means to assertain the quality of products, which leads to a
no-trade paradox \cite{Ak70}. When the customers are better
equipped, optimal quality emerges \cite{Ho52,Sha82,Zha01}. One
of the main issues is to understand under which conditions a
manufacturer should diversify his production. Economics
literature has approached this problem mainly with the help of
utility functions. Several aspects have been studied, among them
optimal quality-based product differentiation~\cite{Mot93}, firm
competition by quality~\cite{Joh03} and by price~\cite{ChaRo89},
the relation between product quality and market
size~\cite{BeWa03}, etc.~(see \cite{Sut91,CaPe05} for a review).

We assume that customers' decisions, while influenced by
perceived properties of the products, are probabilistic in
nature. Using a probabilistic consumer choice framework makes it
possible to avoid utility functions and hence our model can be
understood as an alternative to the usual utility-function
approach. For other alternatives, known as models of discrete or
probabilistic choice, which still use utility theory and yet
they are probabilistic see~\cite{McF80,Cu82,AP92}. In our work
we take the point of view of a monopolistic vendor faced to
consumers deciding to buy one of his products according to their
perception of its quality. The resulting complex complex system
with one vendor, several product variants, and many
heterogeneous buyers, is investigated by numerical techniques.

The paper is organized as follows. In section~\ref{sec:one_var}
we introduce our framework and determine the optimal quality of
a~single product proposed to homogeneous or heterogeneous
customers. In section~\ref{sec:multiple} we examine the
conditions under which a vendor should segment the market by
manufacturing several products of different quality. In
section~\ref{sec:price} we allow the vendor to optimize the
price as well. We leave to the appendices a deeper discussion of
our assumptions and more technical results on the economics of
spamming.

\section{Single product}
\label{sec:one_var}
We assume that the only difference between products lies in
their quality $Q\ge0$ which is therefore the main quantity of
interest here.\footnote{One can also build a model starting from
the tastes of the buyers as in~\cite{MZ07}.}
With a suitable choice of units, one can write the profit of a
vendor per item sold as $1-F(Q)$ where $F$ is an increasing
function and $F(1)=1$. For the sake of simplicity, we take
$F(Q)=Q$; $F(Q)\propto Q^2$ is also found in the literature but
does not alter qualitatively our results. While $Q$ could in
principle be greater than one, a vendor would never choose it,
hence our analysis is restricted to $Q\in[0,1]$.

\begin{figure}
\centering
\includegraphics[scale=0.27]{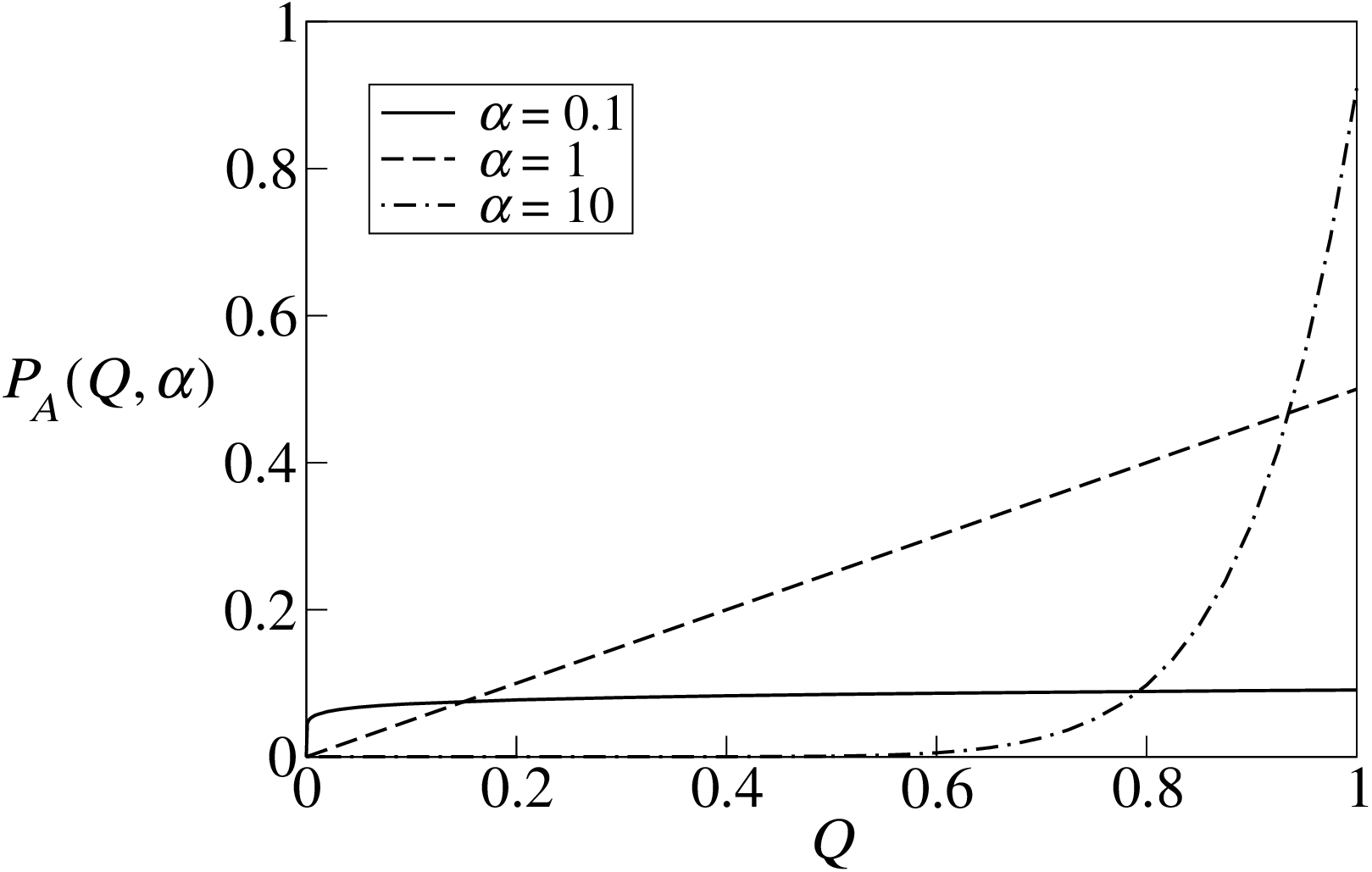}
\caption{The acceptance probability $P_A(Q,\alpha)$ for various
values of the acceptance parameter $\alpha$.}
\label{fig:P_A}
\end{figure}

We assume that a customer buys a product of quality $Q$ with
acceptance probability $P_A(Q)$. While there are many possible
choices, e.g. those of Refs.~\cite{Sut91,Schm86} or piece-wise
linear functions as in Ref.~\cite{Zha01}, we shall mainly use
\begin{equation}
\label{P_A}
P_A(Q,\alpha)=\big(1-\tfrac1{\alpha+1}\big)\,Q^{\alpha}\qquad
(\alpha\geq0)
\end{equation}
where $\alpha>0$ is the acceptance parameter: for small
$\alpha$, $P_A$ is mostly flat, resulting in a lack of quality
discrimination; as $\alpha$ grows, the core of $P_A$ shifts
towards higher quality, which reflects enhanced discrimination
abilities (see Fig.~\ref{fig:P_A}). We will use the shorthands
``ignorant'' for buyers with a small $\alpha$ and ``informed''
for those with a large $\alpha$; an ignorant buyer is quite
likely to reject even a perfect product. Since for $\alpha>0$
and $Q\in[0;1]$ is $P_A(Q,\alpha)<1$, by Eq.~\req{P_A} we
implicitly assume that for the considered product there are
substitutes which can satisfy needs of consumers.

If there are $N$ buyers with acceptance parameters $\alpha_i$
($i=1,\dots,N$), faced with a single product of quality $Q$, the
vendor's expected profit $X$ is
\begin{equation}
\label{X-general}
X(Q)=(1-Q)\sum_{i=1}^N P_A(Q,\alpha_i)-Z.
\end{equation}
where $Z$ represents the fixed part of production costs due, for
instance, to the initial investment needed to setup the
manufacturing plant. Assuming that $N$ is large, the
fluctuations of $X$ can be neglected. The structure of this
expression is similar to the profit function introduced
in~\cite{Sut91}. Since $X'(0)>0$, $X'(1)<0$, and $X(Q)$ is a
continuous function, there is at least one $Q$ maximizing $X$ in
$(0,1)$.

In the following we take the point of view of the vendor and
hence optimize his expected profit $X$.

\subsection{Homogeneous population}
\begin{figure}
\centering
\includegraphics[scale=0.27]{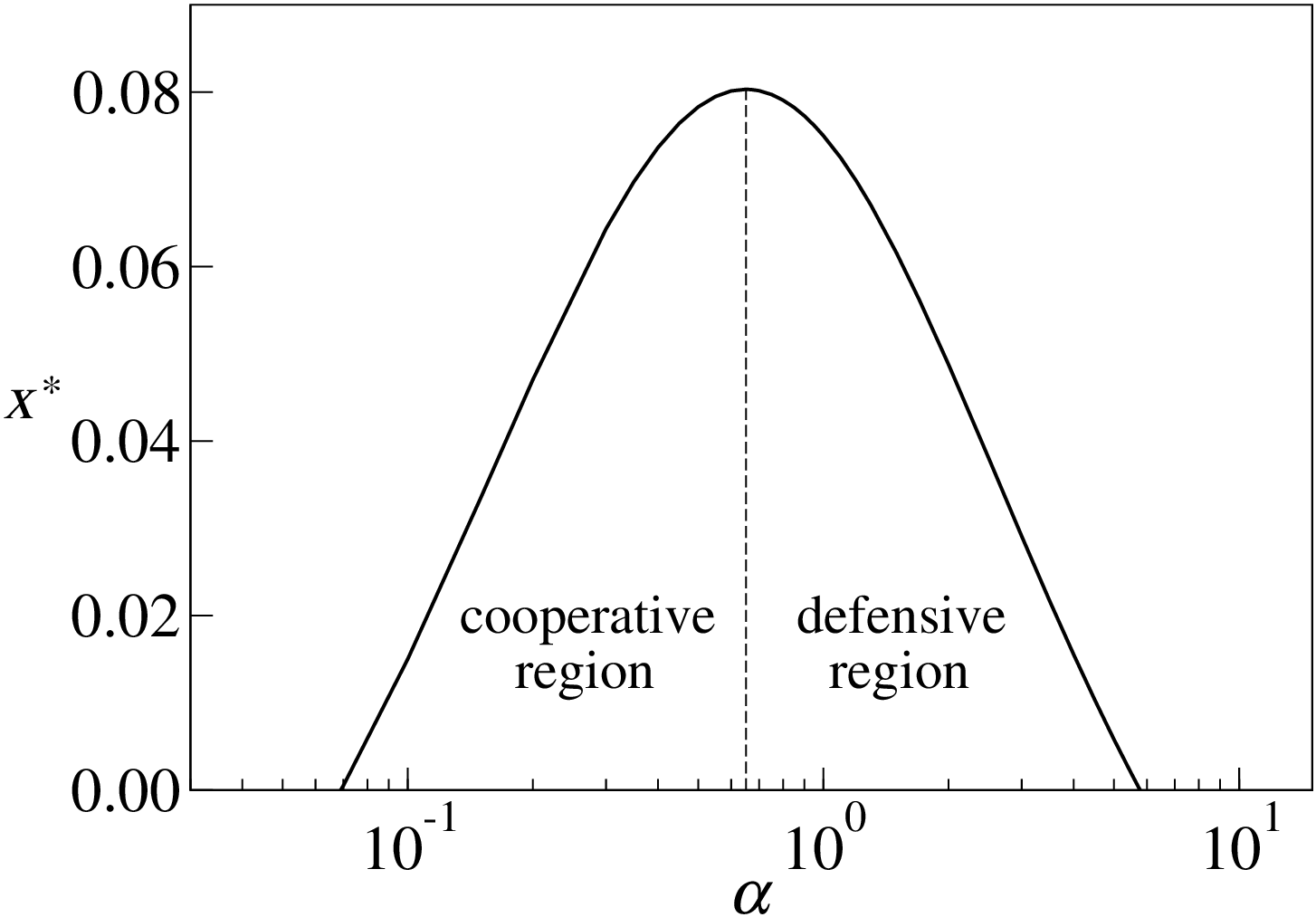}
\caption{Optimal vendor's profit per buyer $x^*$ as a function
of $\alpha$ for $z=0.05$.}
\label{fig:Xopt_k}
\end{figure}

When there is only one type of buyers, the expected profit
simplifies to $X(Q)=N(1-Q)P_A(Q,\alpha)-Z$ which reaches its
maximum at
\begin{equation}
\label{Qopt-1-1}
Q^*(\alpha)=\frac{\alpha}{\alpha+1}.
\end{equation}
Expectedly, $Q^*(\alpha)$ increases when the buyers have a
shar\-per eye. The total optimal profit reads
\begin{equation}
\label{Xmax-1-1}
X^*(\alpha)=N\frac{\alpha^{\alpha+1}}{(\alpha+1)^{\alpha+2}}-Z.
\end{equation}
In Fig.~\ref{fig:Xopt_k} we report the expected optimal profit
per customer $x^*(\alpha):=X^*(\alpha)/N$  as a function of
$\alpha$ for $z:=Z/N=0.05$. When $z>0$, a vendor only makes a
profit when the quality is not too high or too low. Accordingly
$x^*(\alpha)$ has a maximum at $\alpha_0\approx0.65$. Therefore,
if the vendor cannot easily change $Q$, he should target a
population with $Q^*$, or strive to modify the abilities of his
prospective customers to detect quality, thereby increasing his
profit. When $0\leq\alpha<\alpha_0)$, both the consumers and the
vendor benefit from an increase in $Q$; we shall call it the
cooperative region. Reversely, when $\alpha>\alpha_0$ the
vendor suffers from excessive quality detection abilities of
his customers; he could try a confusing marketing campain or
rebranding so as to lower their abilities---this is the
defensive region. A similar behaviour has been observed
in~\cite{Zha05}. In our case, the fact that the cooperative
region is much smaller than the defensive region is a
consequence of the shape of $P_A$. For instance, when the
prefactor in $P_A(Q)$ changes from $1-(\alpha+1)^{-1}$ to
$1-(\alpha+1)^{-1/3}$, the size of the cooperative region
increases significantly.

\subsection{Heterogeneous buyers}
\begin{figure}
\centering
\includegraphics[scale=0.27]{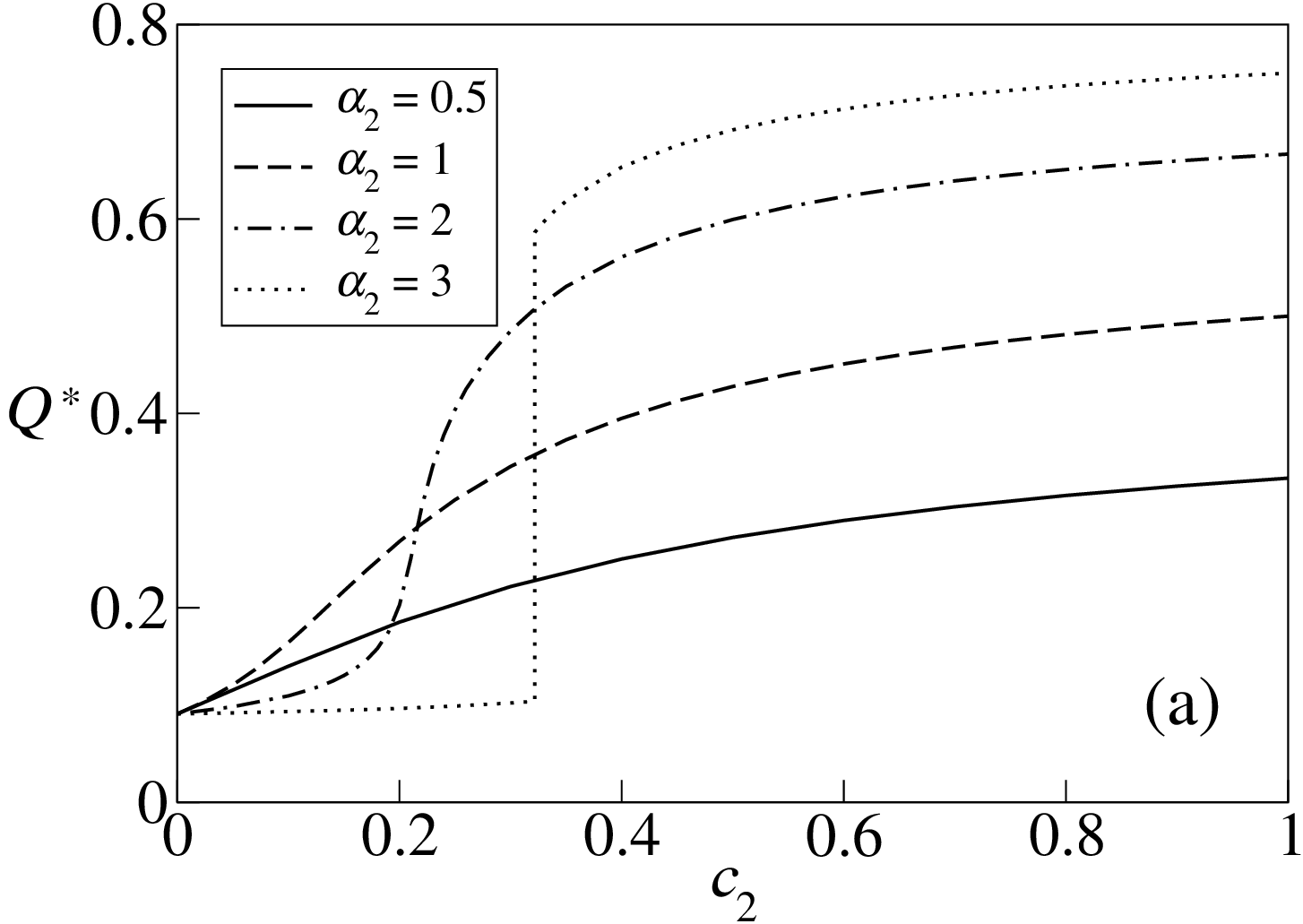}\qquad
\includegraphics[scale=0.27]{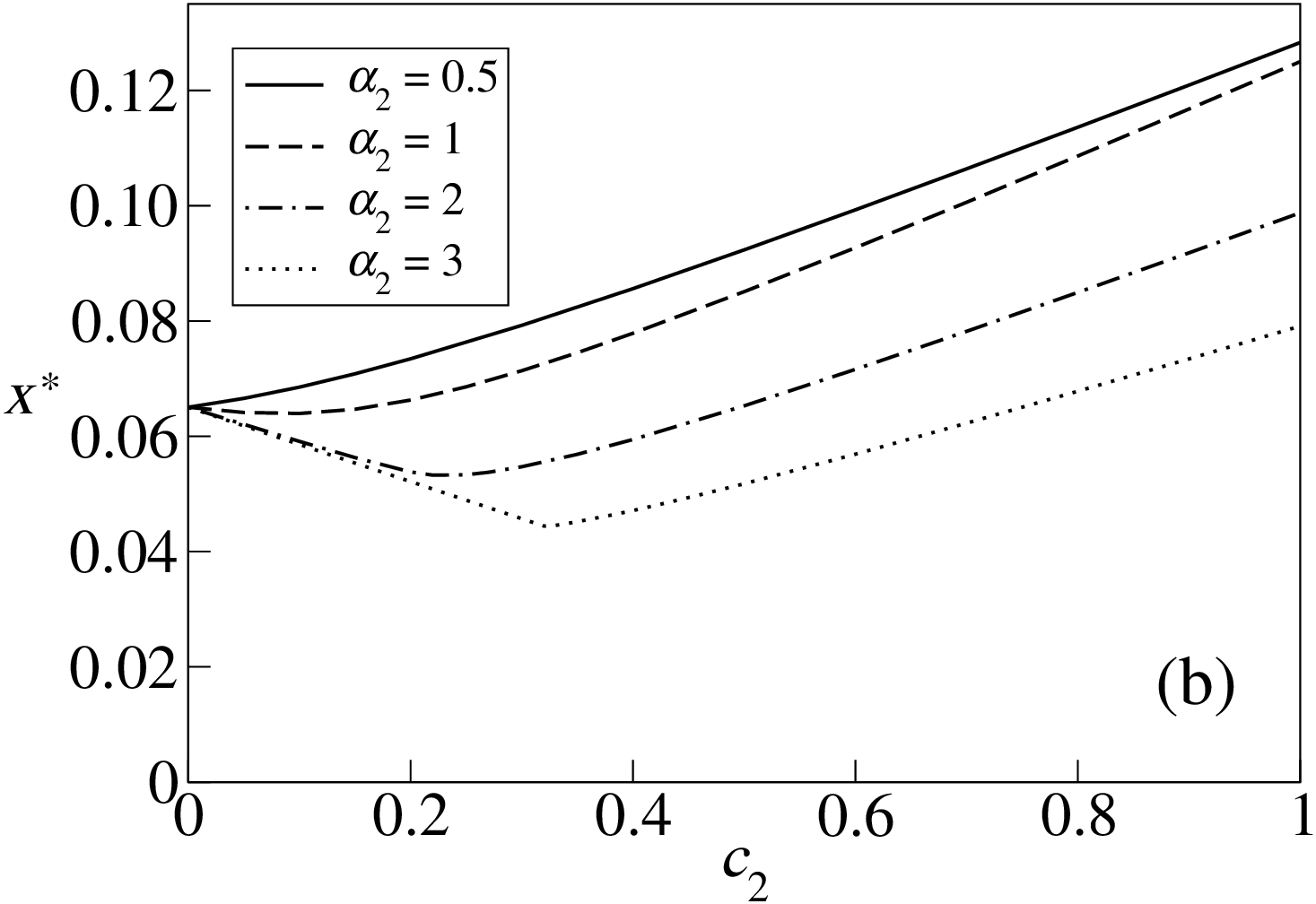}
\caption{Optimal product quality $Q^*$ (left) and vendor's
optimal  profit $x^*$ (right) versus proportion $c_2$ for
various values of $\alpha_2$; $\alpha_1=0.1$, $z=0.01$.}
\label{fig:Qopt-1-2}
\end{figure}

Heterogeneity brings in more surprises. Let us split the
population into two groups, group $i=1,2$ consisting of $N_i$
buyers with acceptance parameter $\alpha_i$; the proportion of
group $i$ is denoted by $c_i:=N_i/N$. The vendor's expected
profit reads
\begin{equation}
X(Q)=N(1-Q)\big[c_1P_A(Q,\alpha_1)+c_2P_A(Q,\alpha_2)\big]-Z.
\end{equation}
It is not possible to maximize $X$ analytically. The result of
numerical investigations is shown in Fig.~\ref{fig:Qopt-1-2} as
a function of $c_2$ for $\alpha_1=0.1$ (ignorant buyers) and
various choices of $\alpha_2$. As expected, as the proportion of
informed buyers increases, $Q^*$ grows. But a surprising
behaviour is found for instance when  $\alpha_2=3$: at
$c_2\approx0.32$ the optimal quality changes discontinuously.
This is because $X(Q)$ has  two local maxima. While for small
$c_2$ the small-$Q$ peak yields the largest profit, its relative
height decreases as $c_2$ increases; accordingly the
discontinuous transition occurs when the heights of the two
maxima are equal. In Fig.~\ref{fig:Qopt-1-2} we also show the
dependence of the optimal profit per buyer $x^*$ on $c_2$. When
group 2 has $\alpha_2<\alpha_0$ (e.g., $\alpha_2=0.5$), adding
people with more demands regarding quality is beneficial to the
vendor (Eq.~\req{Xmax-1-1}) and $X$ is an increasing function of
$c_2$. By contrast, when $\alpha_2>\alpha_0$ the optimal profit
first decreases as almost nobody of group 2 will buy anything
and does so as long as group 2 has less influence on $Q^*$ than
group 1. Then group 2 supercedes group 1 and imposes its quality
demands; the discussion generalises to an arbitrary number of
groups. In other words, when society is too heterogeneous, it is
impossible to satisfy all buyer groups with one product.

\begin{figure}
\centering
\includegraphics[scale=0.27]{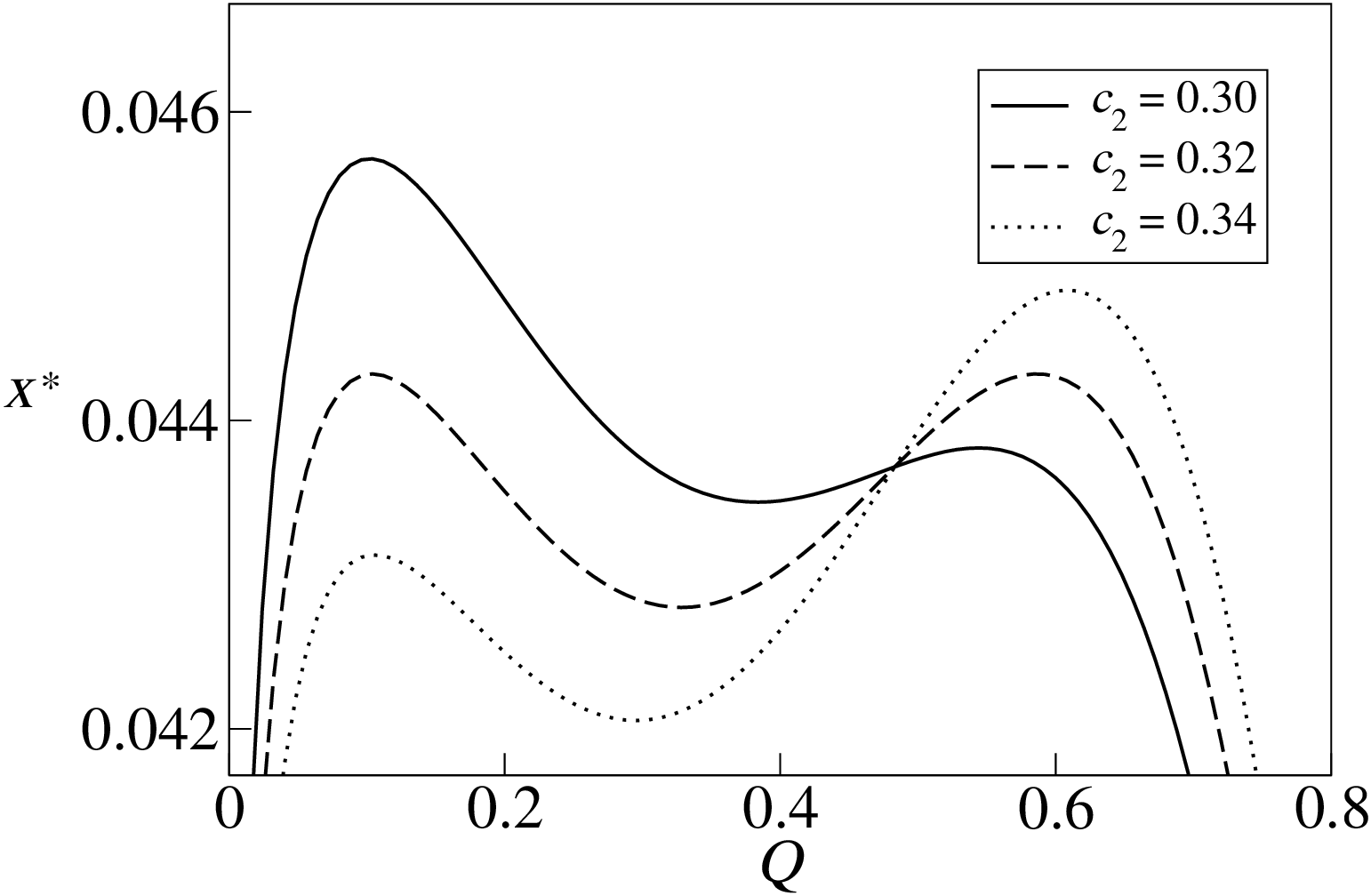}
\caption{Optimal profit per buyer $x^*$ as a function of quality
$Q$: one product, two groups of buyers ($\alpha_1=0.1$,
$\alpha_2=3.0$). As $c_2$ increases, at $c_2\approx0.32$ the
heights of the maxima are equal and a discontinuous change of
the optimal quality occurs.}
\label{fig:maxima}
\end{figure}

\section{Multiple products}
\label{sec:multiple}
Now we assume that the vendor displays $M$ product variants of
different quality, at equal prices for the sake of simplicity,
and that each buyer buys at most one item. A purchase is a
two-step process, as a shopper has also to decide on a variant.
The choice is also assumed to be probabilistic: variant
$m=1,\dots,M$ is chosen according to
\begin{equation}
\label{P_S}
P_S(m\vert\vek{Q},\sigma)=\frac{Q_m^\sigma}{\sum_{m'=1}^M
Q_{m'}^\sigma}.
\end{equation}
Here $\sigma\in[0,\infty)$ quantifies the selection ability of
a given buyer. When $\sigma$ is large, the buyer almost surely
selects the best variant; on the contrary when $\sigma=0$,
$P_S(m)=1/M$ for all $m$, i.e., the buyer has no discerning
power. Since $P_S$ is normalized, each buyer purchases at most
one item. Similar expressions appear in works on the influence
of advertisement~\cite{Schm86} and non-price
competition~\cite{Sut91}, but other choices of functions would
also be reasonable, such as exponentials as in the Logit
model~\cite{McF74,AP92}. All $Q_{m'}^{\sigma}$ have equal weight
in Eq.~\req{P_S}; section~\ref{sec:proportions} generalizes this
expression in order to take into account the proportions of
displayed items. Finally, a more complete discussion on the
plausibility of $P_S$ is given in Appendix~\ref{app:discussion}.

\begin{figure}
\centering
\includegraphics[scale=0.27]{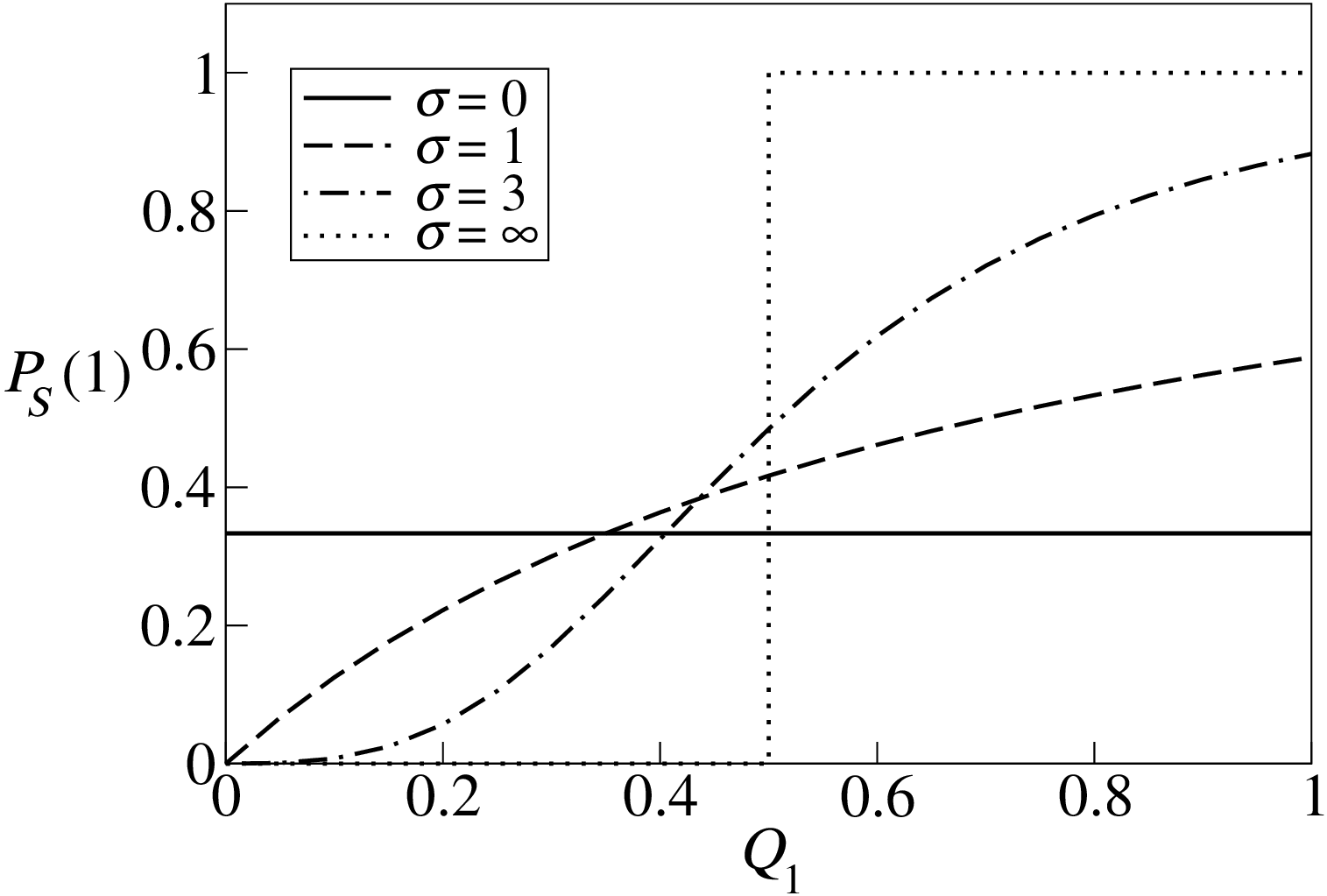}
\caption{The probability to select variant 1 as a function of
its quality $Q_1$ for various values of the selection parameter
$\sigma$. In total three variants are displayed, the
qualities $Q_2=0.5$ and $Q_3=0.2$ are fixed.}
\label{fig:P_S}
\end{figure}

To summarize, the variant $m$ with the quality $Q_m$ is bought
by buyer $i$ with probability
$P_S(m\vert\sigma_i,\vek{Q})P_A(Q_m,\alpha_i)$. As a
consequence, if the vendor displays $M$ variants to $N$ buyers,
his expected profit is
\begin{equation}
\label{X-selection}
X(\vek{Q})=\sum_{m=1}^M (1-Q_m)\bigg(\sum_{i=1}^N
P_S(m\vert\vek{Q},\sigma_i)P_A(Q_m,\alpha_i)\bigg)-MZ.
\end{equation}
This equation can be easily extended to account for special
circumstances. For example, when $Z$ is large, it may be
profitable to produce one variant and achieve quality
differentiation by artificially damaging a fraction of the
production, e.g. by  disabling some features~\cite{McA07}. In
this case two variants with qualities $Q_1>Q_2$ are displayed
but the profit per item sold is only $1-Q_1$ for both of them
and the initial cost is reduced from $2Z$ to $Z$.

\subsection{Quality differentiation}
\label{sec:differentiation}
For the sake of simplicity, we focus on two groups of customers
consisting of $N_i$ members with acceptance parameter $\alpha_i$
and selection power $\sigma_i$ ($i=1,2$). The question is
whether the vendor should display one or two products. In our
framework, the answer is entirely determined by the respective
optimal profit of each possibility, denoted by $X_1^*(Q)$ and
$X_2^*(Q_1,Q_2)$.

\begin{figure}
\centering
\includegraphics[scale=0.27]{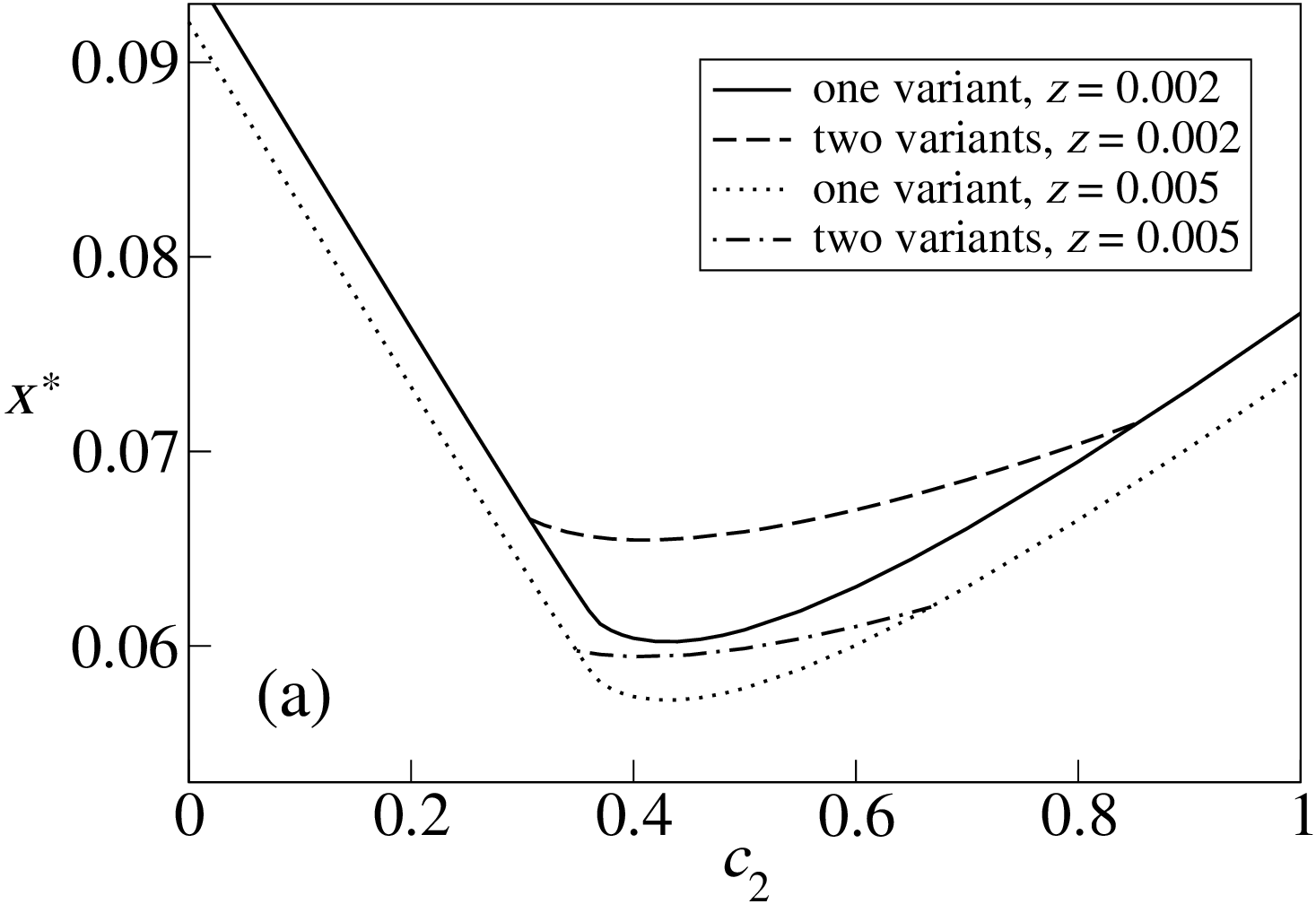}\qquad
\includegraphics[scale=0.27]{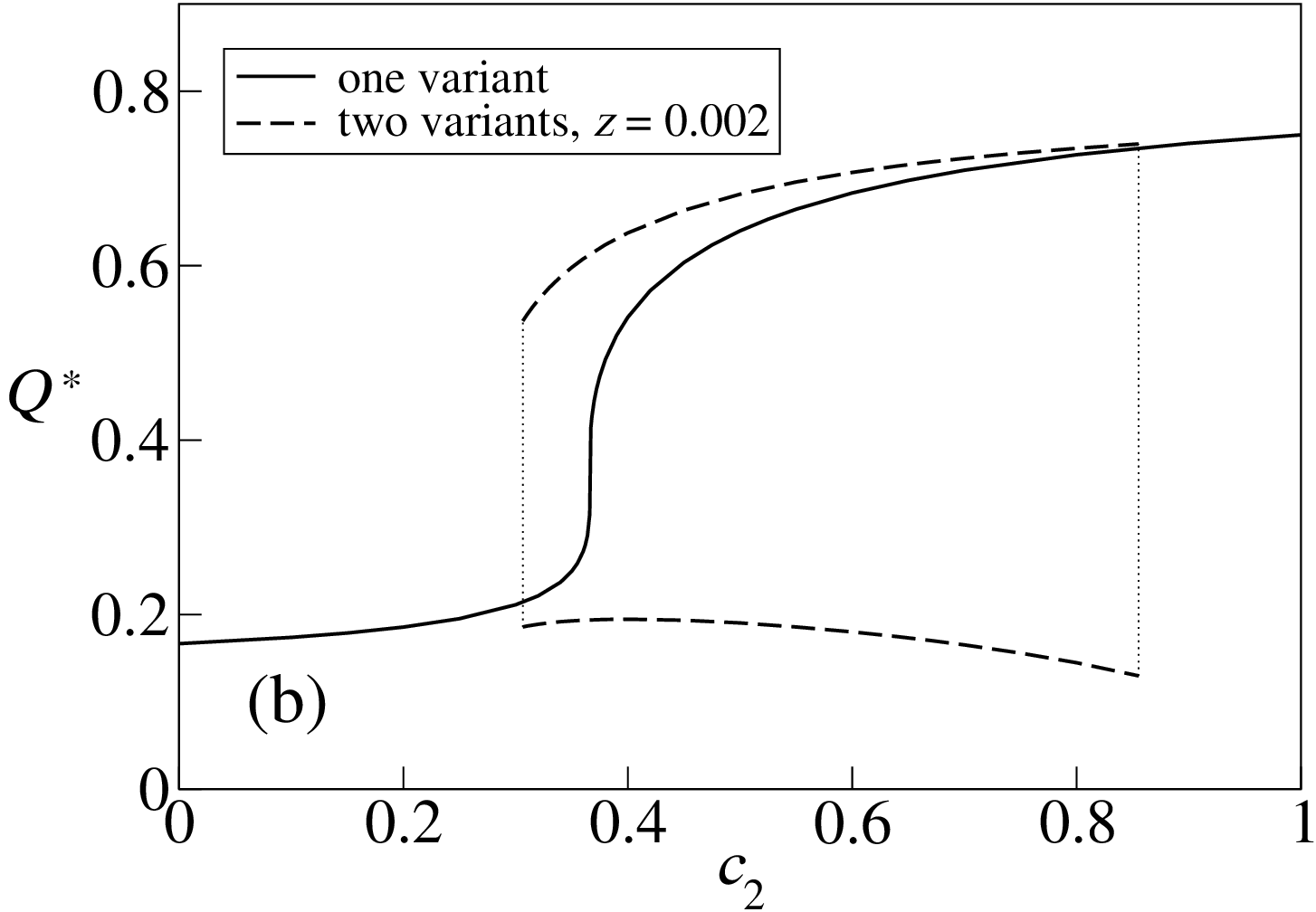}
\caption{a) Optimal profits per buyer as a function of
proportion $c_2$; two-variant profits $x_2^*$ (broken and
dashdot lines) are shown only when quality differentiation
occurs. b) Optimal quality as a function of proportion $c_2$,
curves for the differentiated qualities $Q_1^*$ and $Q_2^*$ are
only shown when $x_2^*>x_1^*$. Values of parameters are
$\alpha_1=0.2$, $\alpha_2=3.0$, $\sigma_1=0.5$, and
$\sigma_2=3.0$.}
\label{fig:differentiation}
\end{figure}

Since manufacturing two products requires twice as much initial
investment (by hypothesis), the region in which $X_2^*>X_1^*$
shrinks when $z$ increases. This appears clearly in
Fig.~\ref{fig:differentiation} where we plot the optimal profits
versus $c_2=N_2/N$ for two values of $z$. In addition, when
$X_2^*>X_1^*$, the two optimal qualities $Q_1^*$ and $Q_2^*$
differ significantly. Quite clearly, the lower quality targets
the group of ignorant buyers while the higher quality is for
informed buyers. Remarkably, when $c_2>0.68$, the lower optimal
quality is even smaller than the optimal quality
$\alpha_1/(\alpha_1+1)=1/6$ corresponding to a homogeneous
population of ignorant customers. This downward distortion in a
situation of a monopolistic vendor is also reported
in~\cite{MuRo78}; it is optimal as it reduces the substitution
possibilities of higher-value (or numerous enough) customers.
The benefits of low quality variants in market competition are
discussed in detail in~\cite{Joh03}.

\subsection{How many variants?}
\label{sec:regions}
\begin{figure}
\centering
\includegraphics[scale=0.27]{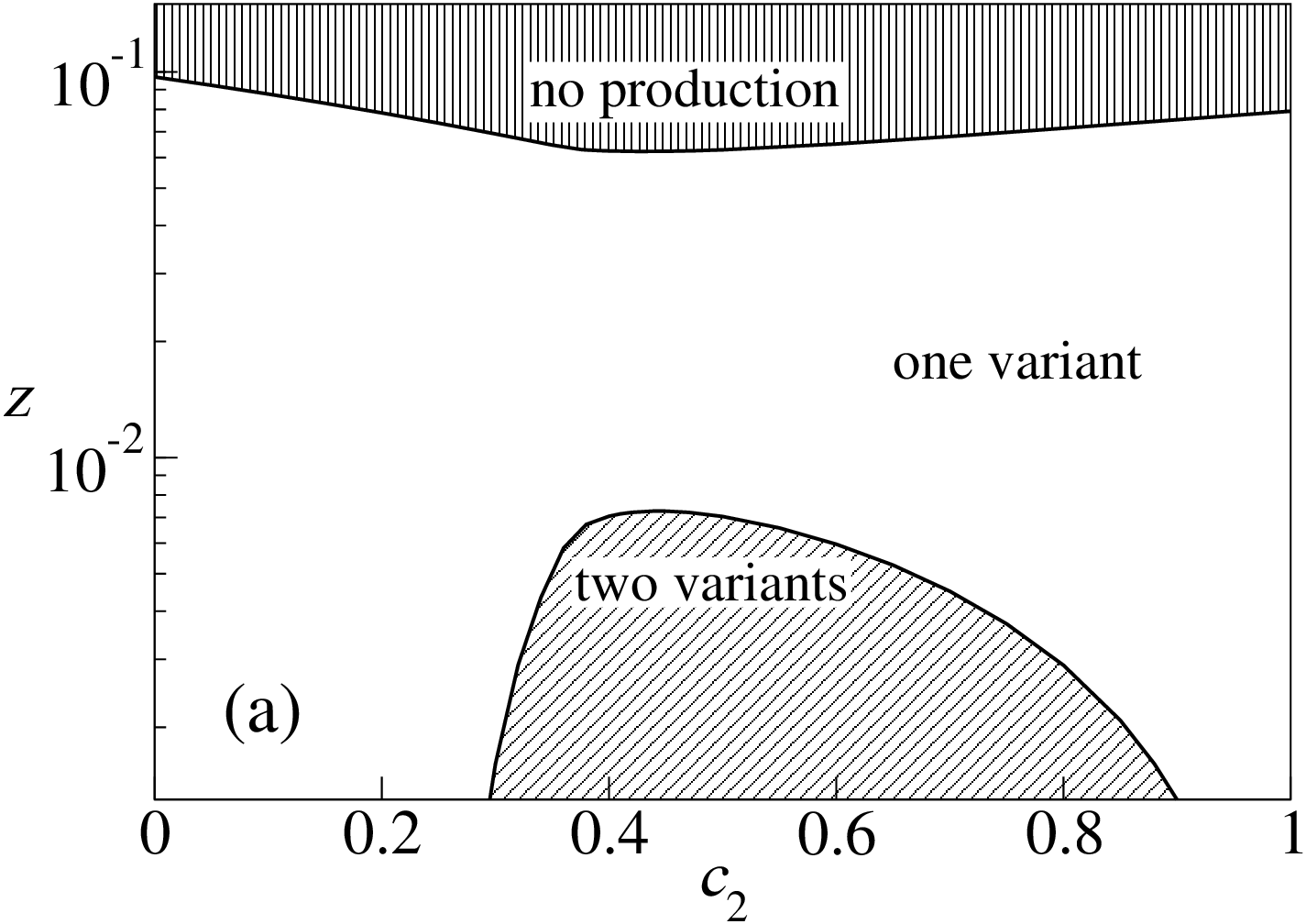}\qquad
\includegraphics[scale=0.27]{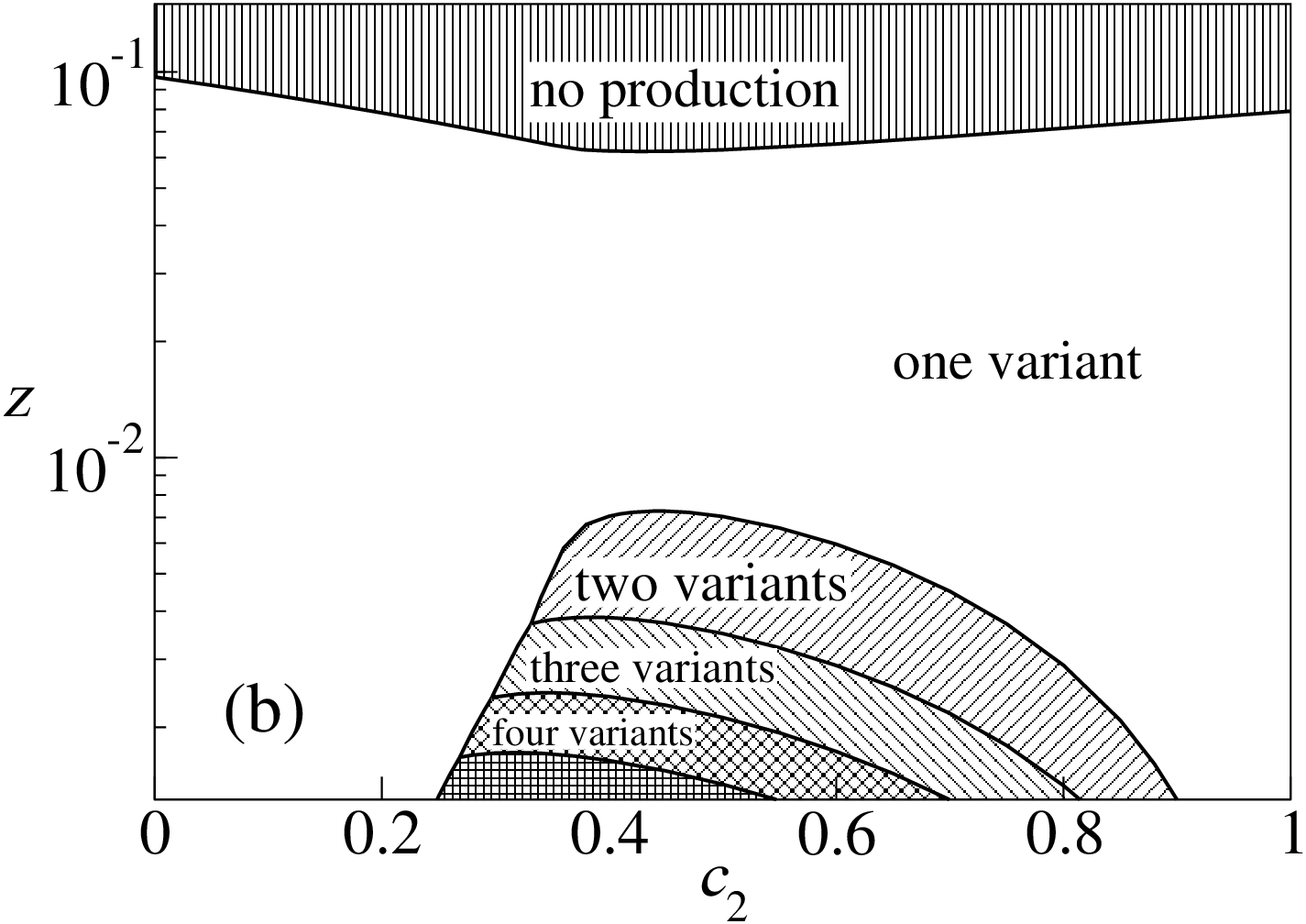}
\caption{Phase space $(c_2,z)$ of optimal production: only
$0,1,2$ variants are allowed (a) and without constraints on the
maximal number of variants (b). Same parameters as in
Fig.~\ref{fig:differentiation}.}
\label{fig:z_trans}
\end{figure}

The phase space $(c_2,z)$ of optimal production when at most two
variants are allowed is reported in Fig.~\ref{fig:z_trans}a.
At intermediate values of $c_2$, product differentiation exists
if $z$ is small enough (see also \cite{MZ07}). Can a further
decrease of $z$ differentiate further the production?

Figure \ref{fig:z_trans}b reveals the Russian-dolls structure of
product differentiation. Let us consider the case $M=3$: when
$X_3^*>X_2^*$, in fact only two products are really different,
i.e. $Q_1^*=Q_2^*<Q_3^*$. In other words, it pays to duplicate
the low quality variant. This is because it decreases the
likelihood that an ignorant buyer selects the high quality
variant, while informed buyers, thanks to their high selection
parameter, are still able to pick the premium variant. This
mechanism is at work for a generic $M$: when $z$ is small, for
the vendor it may be optimal to display $M-1$ low-quality
variants with identical qualities and one premium variant.

\begin{figure}
\centering
\includegraphics[scale=0.27]{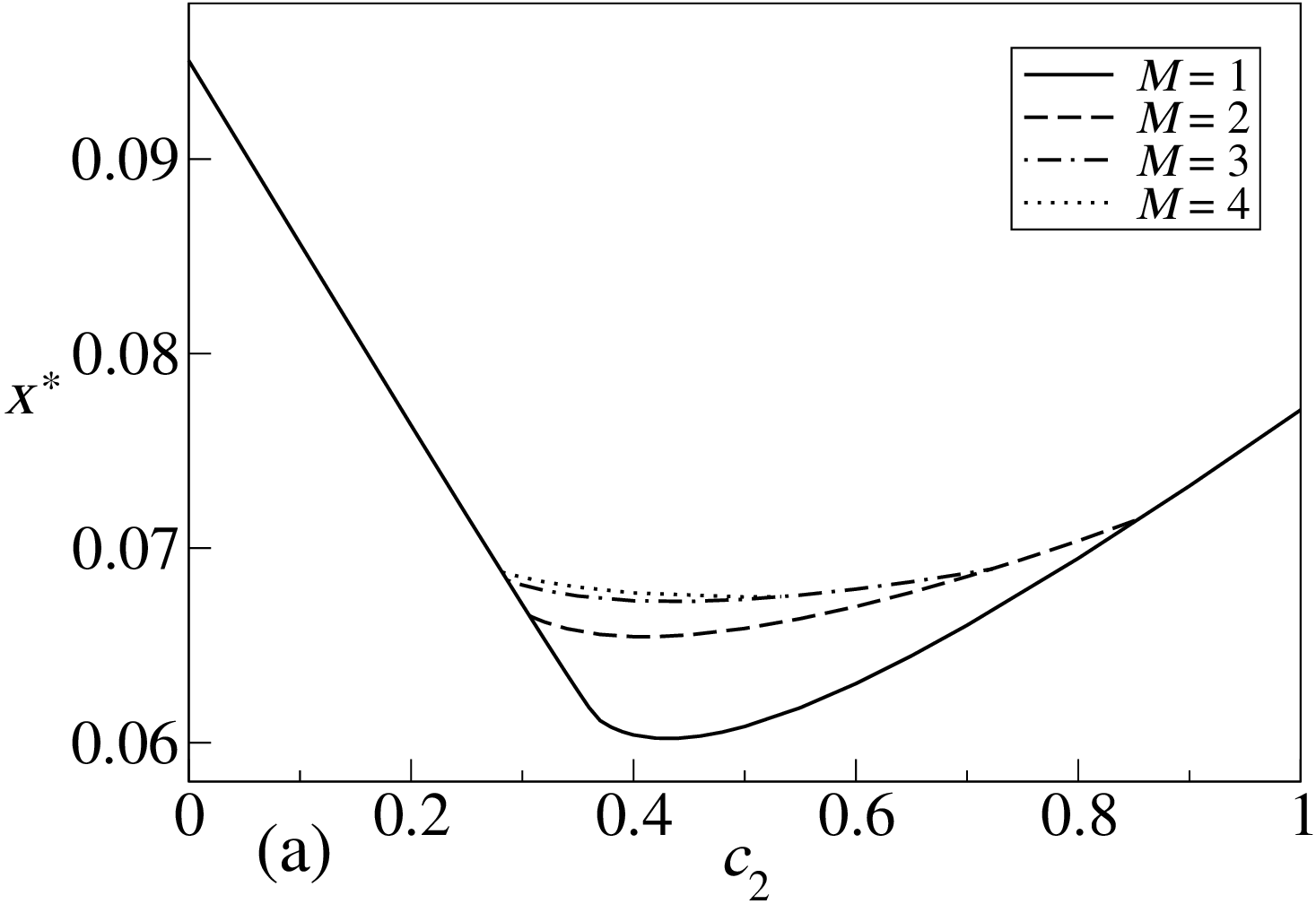}\qquad
\includegraphics[scale=0.27]{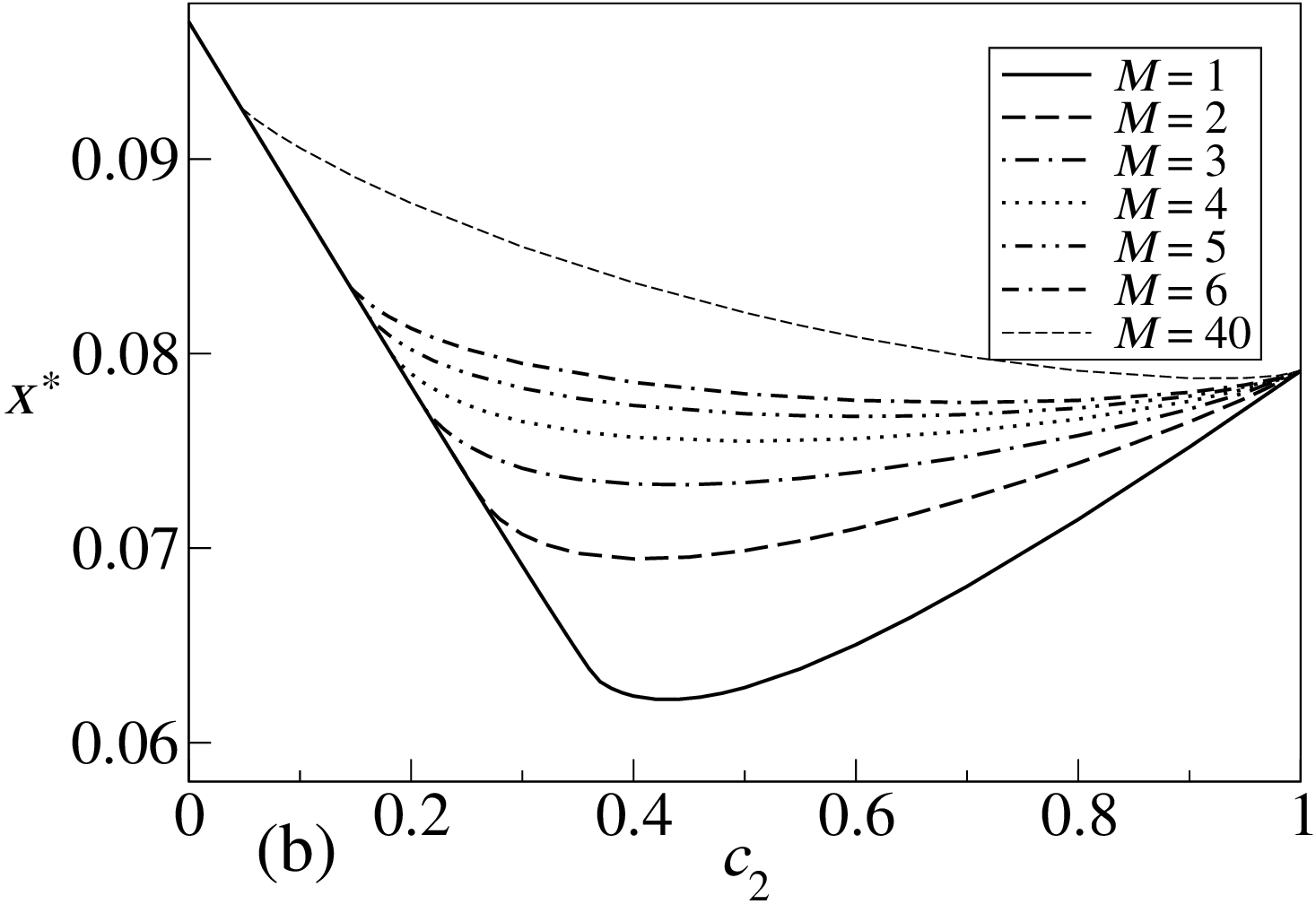}
\caption{Optimal profit for $M$ displayed variants when
$z=0.002$~(a) and when $z=0$~(b). Same parameters as in
Fig.~\ref{fig:differentiation}.}
\label{fig:fine}
\end{figure}

Finally we consider the vendor's expected profit for various
numbers of displayed variants. As can be seen in
Fig.~\ref{fig:fine}a, when $z=0.002$, the additional gain
decreases very fast when $M$ increases and vanishes when $M>4$.
By contrast, when $z=0$, $X^*$ saturates at the much higher
$M=40$ and then decreases.

\subsection{Biased selection}
\label{sec:proportions}
\begin{figure}
\centering
\includegraphics[scale=0.27]{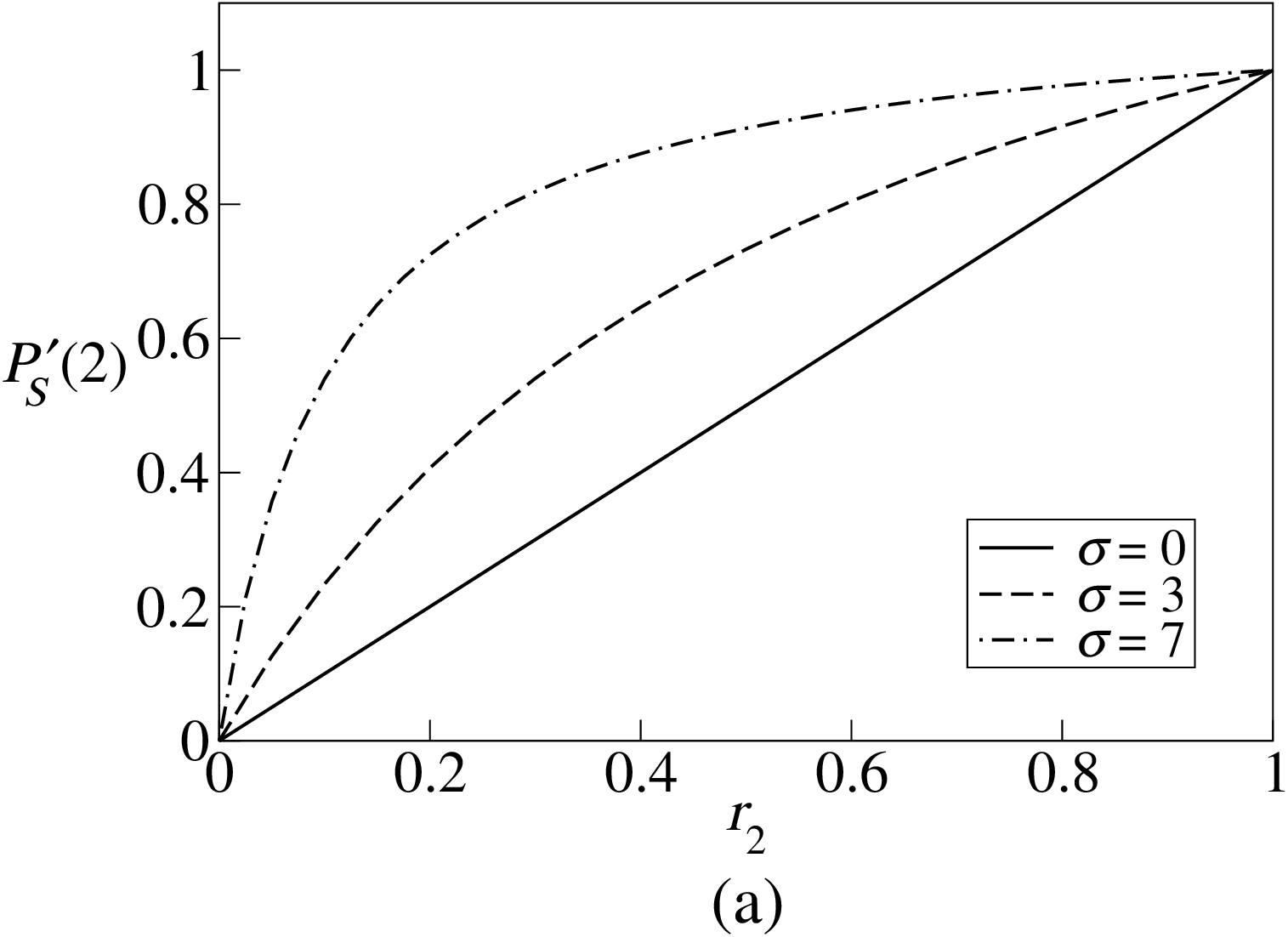}\qquad
\includegraphics[scale=0.27]{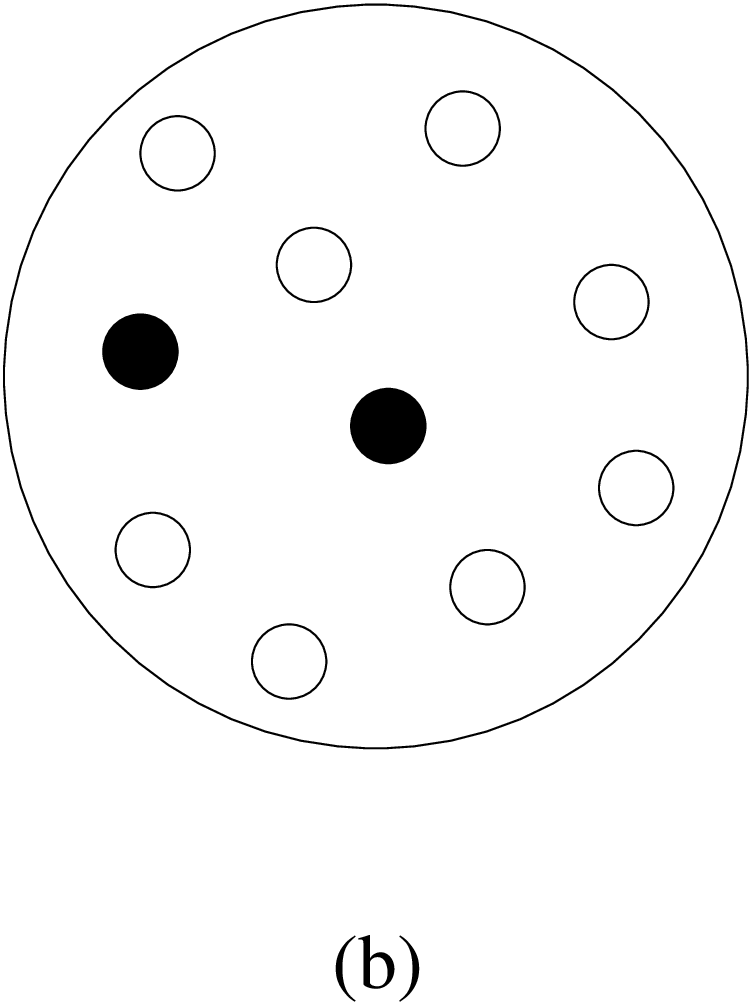}\qquad
\includegraphics[scale=0.27]{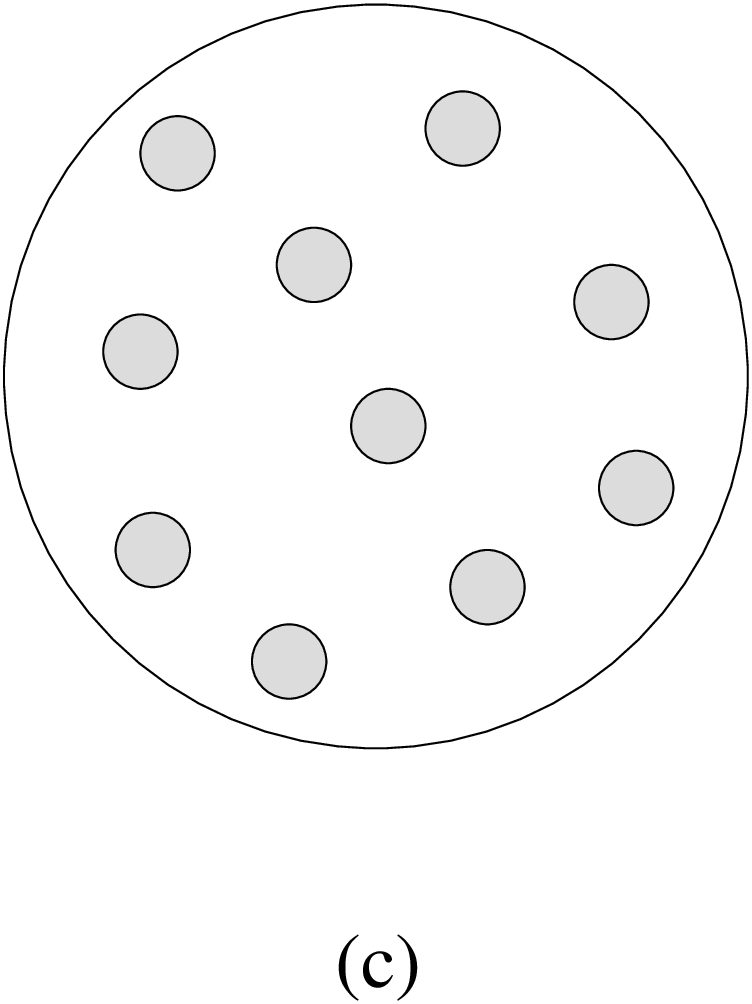}
\caption{a) The probability to select variant 2 in the presence
of two variants with similar qualities ($Q_1=0.5$, $Q_2=0.7$)
for various $\sigma$. b) When qualities are differentiated,
informed buyers are able to select the premium variant (dark
symbols) even when its proportion is small; ignorant buyers
select mostly the low quality variant (white symbols). c) When
only one quality is displayed, the vendor has to compromise
between the two groups and a mediocre variant is optimal (grey
symbols).}
\label{fig:illustrations}
\end{figure}

In Eq.~\req{P_S} we implicitly assume equal standing of the
available variants, which is often not the case in practice.
This suggests to introduce variant weights $r_m$
($\sum_{m=1}^M r_m=1$) in the selection probability $P_S$,
taking into account for instance the effective visibility of
each product due to advertisement or display position in shops.
Eq.~\req{P_S} generalizes to
\begin{equation}
\label{P_S-prop}
P_S'(m\vert\vek{Q},\vek{r},\sigma)=
\frac{r_mQ_m^\sigma}{\sum_{m'=1}^M r_{m'}Q_{m'}^\sigma}.
\end{equation}
An example of the interplay between $r_m$ and $\sigma$ is shown
in~Fig.~\ref{fig:illustrations}: better equipped customers are
able to pick the better product even when its effective
proportion is small.

\begin{figure}
\centering
\includegraphics[scale=0.27]{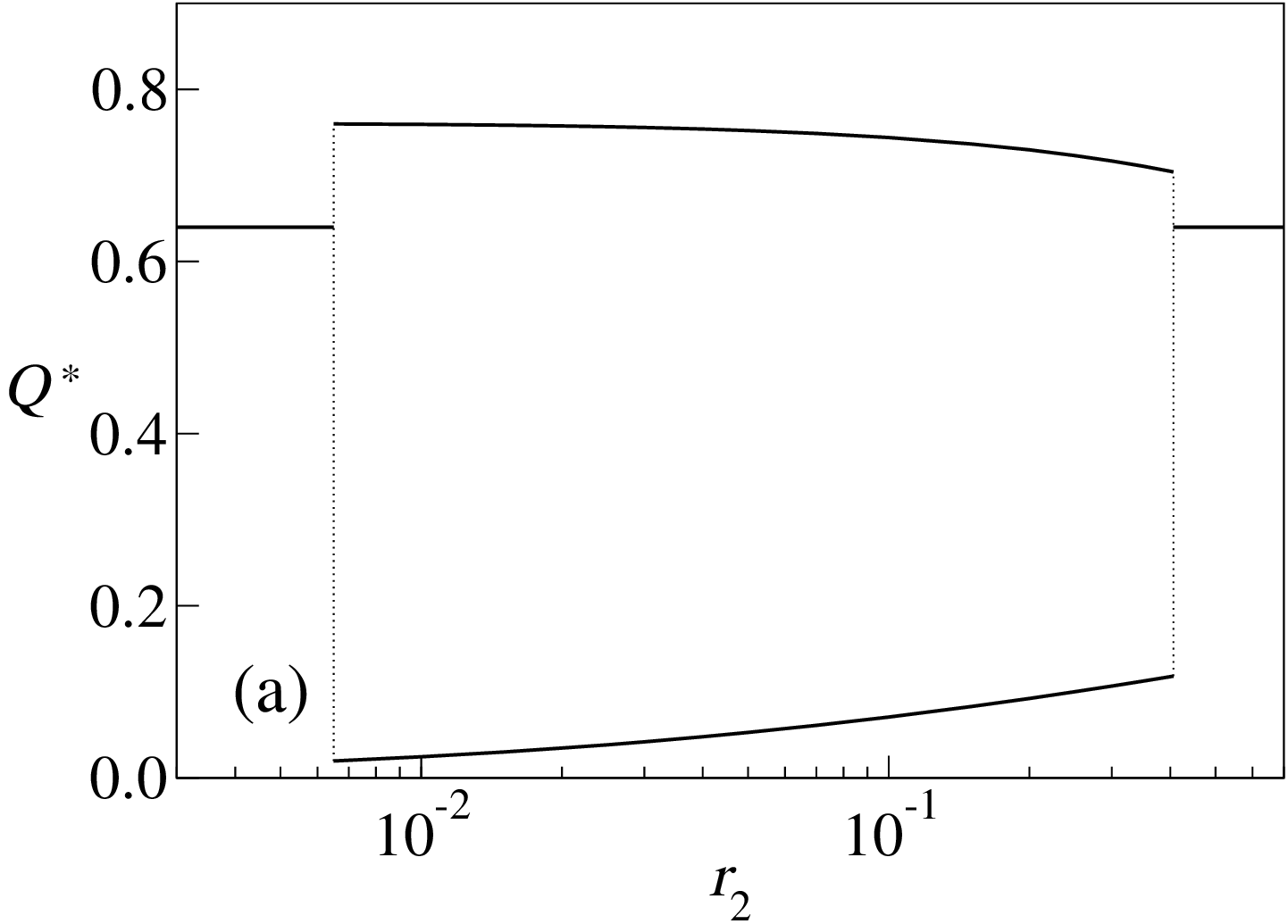}\qquad
\includegraphics[scale=0.27]{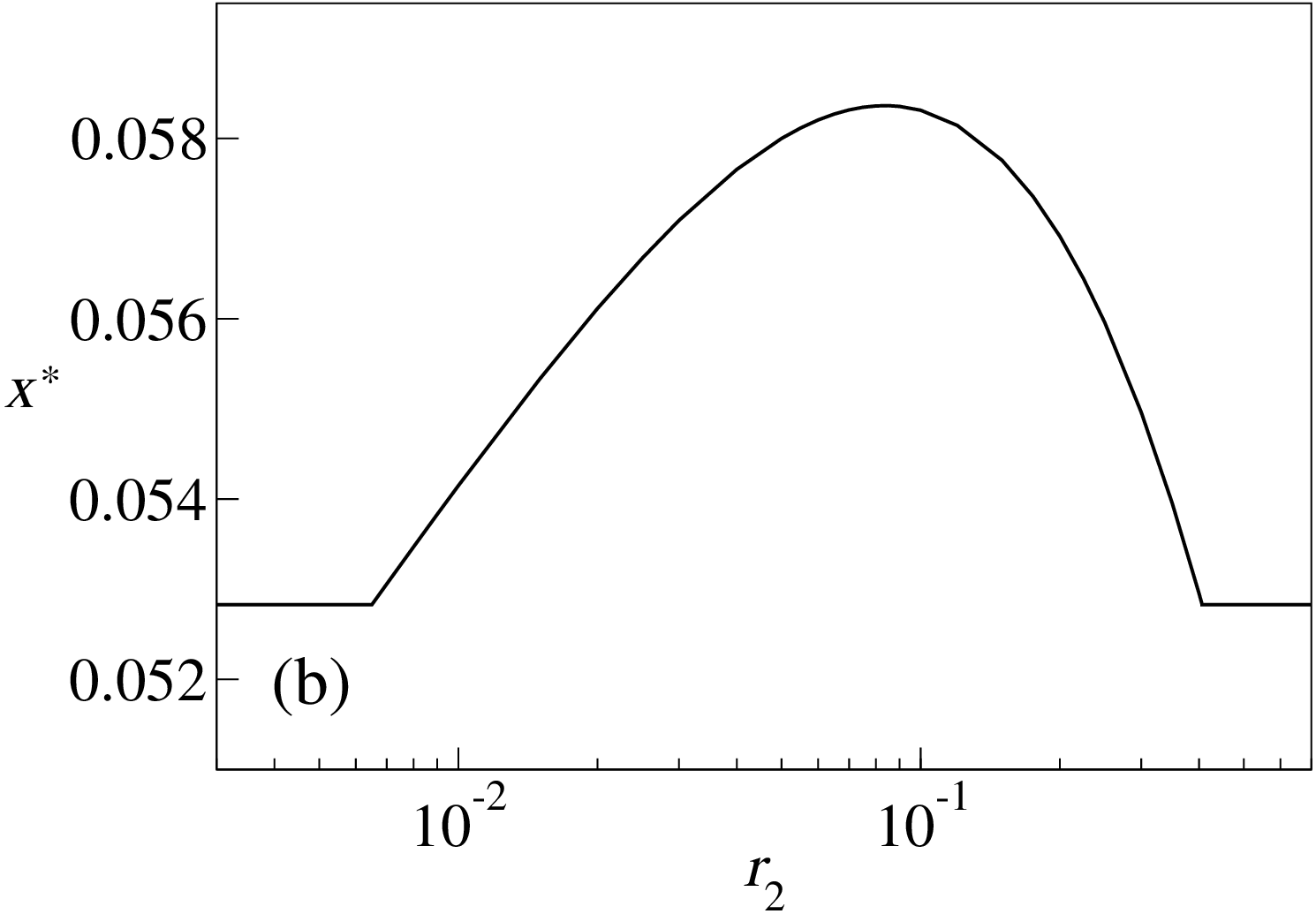}
\caption{Optimal qualities (a) and the optimal profit (b) versus
proportion of the premium variant $r_2$ ($\alpha_1=0.2$,
$\alpha_2=3$, $\sigma_1=0.2$, $\sigma_2=2$, $z=0.01$, $c_2=0.5$,
$M=2$).}
\label{fig:proportions}
\end{figure}

To study the effects of the proposed generalization we use once
again two groups of customers and choose the parameters so as to
set the system in the quality differentiation region. Results of
numerical optimization of the optimal profit are reported in
Fig.~\ref{fig:proportions}, $r_2$ denotes the proportion of the
premium variant. Differentiation occurs in a limited range of
$r_2$: when $r_2$ is either too small or too large, buyers
effectively notice only one variant and it is preferable for the
vendor to produce only that one. In addition, $x^*$ has a
maximum at $r_2\approx0.08$, which comes from hiding the high
quality variant to ignorant buyers while keeping it accessible
to informed buyers.

\section{Price}
\label{sec:price}
Let us now consider the price as a free parameter and
investigate how the vendor should fix it optimally. Denoting the
price by $p$, the profit per item is $p-Q$ which means that the
maximum quality is $Q_{\text{max}}=p$. In particular, if the
vendor wishes to produce a better product that $Q_{\text{max}}$,
the price needs to be increased. The acceptance probability
generalizes to $(\alpha\geq0,\ p\in[Q,\alpha+1])$
\begin{equation}
\label{P_A-price}
P_A(Q,p,\alpha)=\bigg(1-\frac{p}{\alpha+1}\bigg)\,
\bigg(\frac Qp\bigg)^{\alpha}.
\end{equation}
It satisfies two constraints: first, the higher the price, the
smaller the acceptance probability. Second, because of the
$p/(\alpha+1)$ term, the sensitivity towards prices decreases as
sensitivity to quality increases; similarly, quality must be
judged with respect to price, hence the $(Q/p)^{\alpha}$ term.
The discussion of the previous sections corresponds to $p=1$.

\begin{figure}
\centering
\includegraphics[scale=0.27]{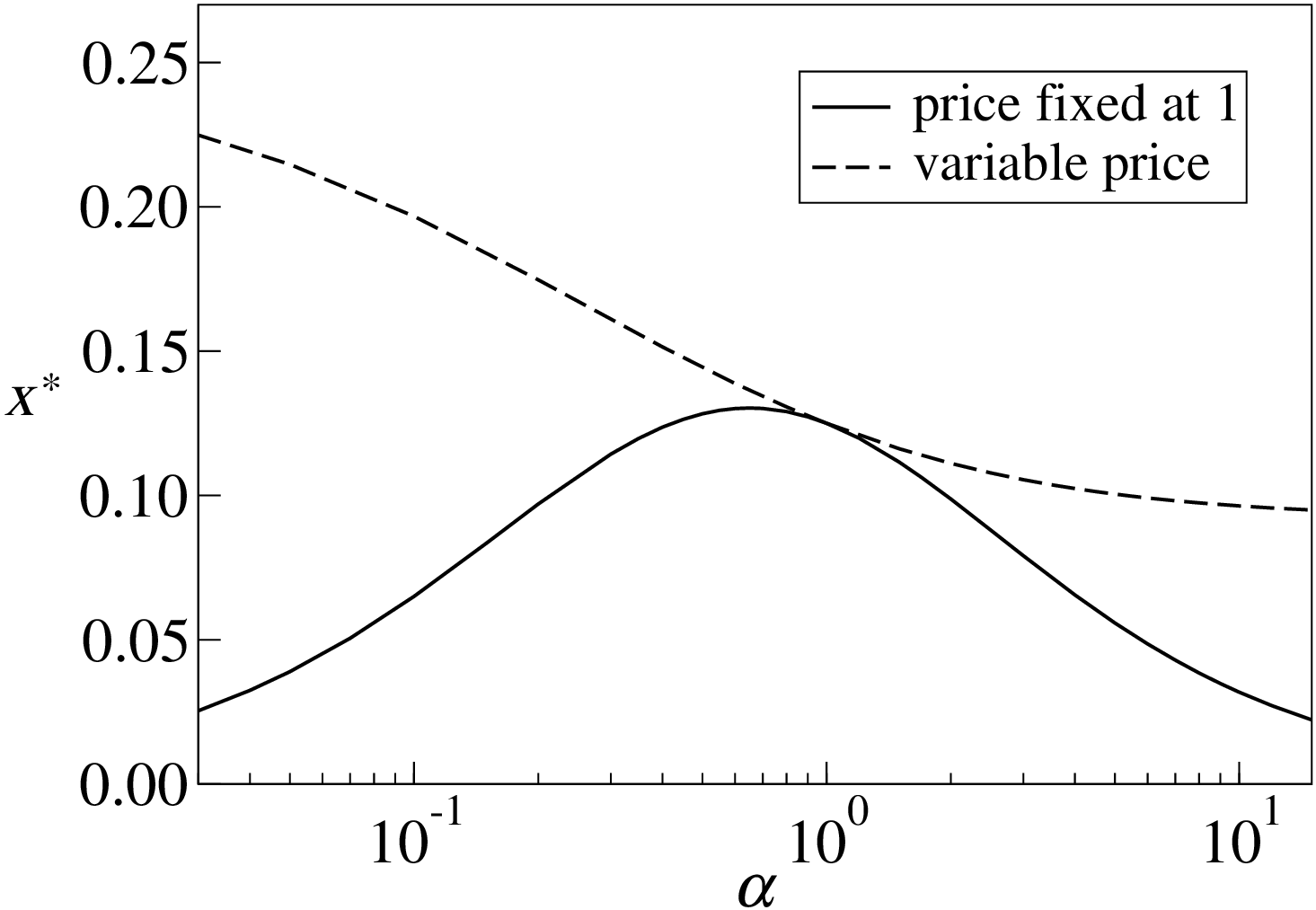}
\caption{Optimal profit $x^*$  vs acceptance parameter $\alpha$:
fixed price (solid line, the same curve as in
Fig.~\ref{fig:Xopt_k}) and variable price (dashed line), $z=0$.}
\label{fig:prices}
\end{figure}

We restrict our analysis to the simplest case of $N$ identical
buyers and one product. The expected profit reads
\begin{equation*}
X_1(Q,p)=N(p-Q)\bigg(1-\frac{p}{\alpha+1}\bigg)\,
\bigg(\frac Qp\bigg)^{\alpha}
\end{equation*}
with $p\leq\alpha+1$ and $Q\in[0,p]$, it is maximized by
\begin{equation}
\label{opt-price}
Q^*=\frac{\alpha}{2},\quad
p^*=\frac{\alpha+1}{2}.
\end{equation}
Expectedly, the more informed the buyers, the better the
products should be, but the vendor can charge a higher price.
Because $p^*-Q^*=1/2$ is a constant, there is no incentive in
this model for exceptionally high prices for high quality
variants. In Fig.~\ref{fig:prices}, the resulting optimal profit
per buyer $x^*$ is shown together with the optimal profit when
the vendor has fixed the price at $1$. The liberty to set the
price can increase the profit of the vendor quite considerably.
The difference of profit for $\alpha>1$ (informed buyers) is due
to the fact that the vendor is allowed to charge a higher price
for the high quality demanded by the buyers. By contrast, for
$\alpha<1$ the main improvement comes from the fact that
$P_A(Q,p,\alpha)$ does not vanish when $Q\to0$ and $p<1$.

\section{Conclusion}
Due to the complexity of markets and human behaviour, attempts
to propose a theory of the whole are illusory. However, simple
models can bring insight to elementary mechanisms at work in the
real economy. Assuming probabilistic buyer behaviour, we
formalized buyers' abilities, spanning from the zero information
to the perfect information limits. Adopting the vendor's point
of view, we examined the compromise between low quality which
minimizes production costs and high quality which maximizes
sales. In particular, the fact that customers are heterogeneous
forces vendors to diversify their production. In other words,
the large variety of products in free-market economies reflects
in part the information gathering and processing abilities of
customers.

In this work we focused on the basic market phenomena but the
proposed model is versatile enough to represent more complicated
cases. Three important extensions seem particularly worth
further investigation: including explicitely the price in
heterogeneous populations, generalizing the present results to
an arbitrary number of consumer groups, and adding more vendors
and letting them to compete for customers (then it can be
Nash stable that variants with different qualities are provided
by different vendors).

\section{Acknowledgment}
This work was supported by the Swiss National Science Foundation
(project 205120-113842) and in part by the International
Scientific Cooperation and Communication Project of Sichuan
Province in China (Grant No. 2008\-HH0014), the stay of
Linyuan L\"u in Fribourg was supported by SBF (Switzerland)
through project C05.0148 (Physics of Risk). We acknowledge the
early contribution of Paolo Laureti and the work of our
reviewers.

\appendix
\section{Discussion of the model plausibility}
\label{app:discussion}
In order to better understand the need for both selection and
acceptance procedures, it is worthwhile to consider some
alternatives. One possibility to simplify our assumptions is to
keep only the acceptance process with each displayed variant
accepted or not according to the acceptance probability $P_A$.
When $M$ variants with the qualities $Q_1,\dots,Q_M$ are
displayed, the probability $P'$ that a given customer accepts at
least one of them is
\begin{equation}
\label{Paccept}
P'(Q_1,\dots,Q_M)=1-\prod_{a=1}^M \big[1-P_A(Q_a)\big].
\end{equation}
As $M$ increases, $P'$ converges to one. This means that the
vendor can attract the buyers by displaying a large number of
very bad products which is generally not the case. However,
flooding of customers by low quality occurs under some special
circumstances. This \emph{economics of spamming} is briefly
discussed in the next Appendix.

Another approach is to reduce the model to the best product
selection governed by the selection probability $P_S$. Since
this probability is normalized to one, when it is applied alone,
each buyer surely buys one of the displayed variants and
consequently the vendor's profit maximization yields zero
quality. Obviously, such an optimal solution is pathological.
One could eventually consider replacing the unity in the
equation $\sum_{m=1}^M P_S(m\vert\vek{Q},\sigma)=1$ by an
increasing function of the displayed qualities but this is
effectively equivalent to our two step decision process. Another
possibility is to introduce an artificial non-purchase
alternative to Eq.~\req{P_S} with an imposed utility as
in~\cite{AP92} which focuses on price differentiation. We see
that nor the selection from the available variants is sufficient
to model the purchase process.

Finally, the generalization to diverse proportions of displayed
variants, introduced in section~\ref{sec:proportions}, gives an
additional argument. We see that while in the selection step
both quality and proportion play their roles, in the final
acceptance step it is only quality of the selected variant what
matters. Thus these two steps are intrinsically different and
attempts to merge them are artificial.

\section{Economics of spamming}
\label{app:spam}
\begin{figure}
\centering
\includegraphics[scale=0.27]{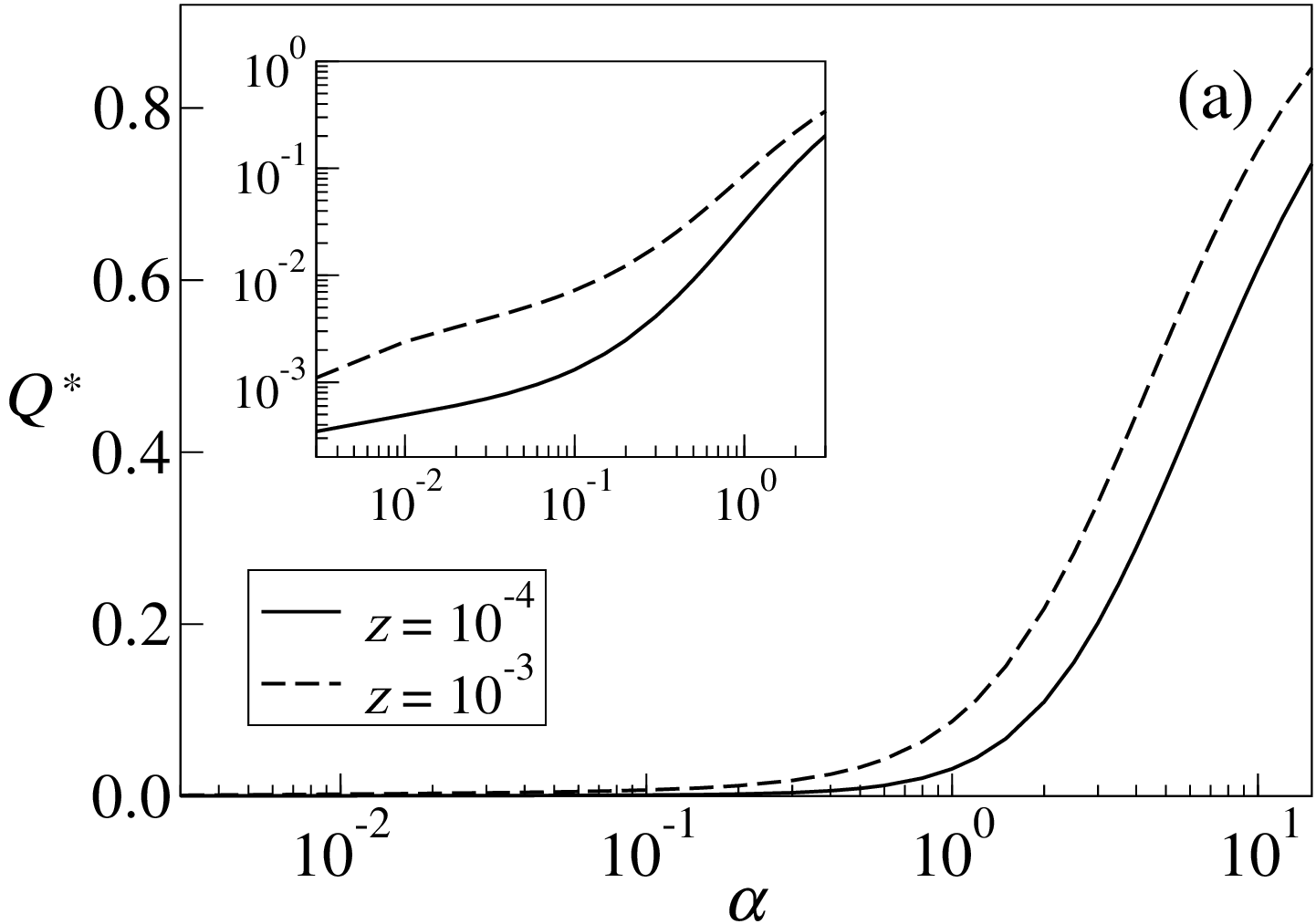}\qquad
\includegraphics[scale=0.27]{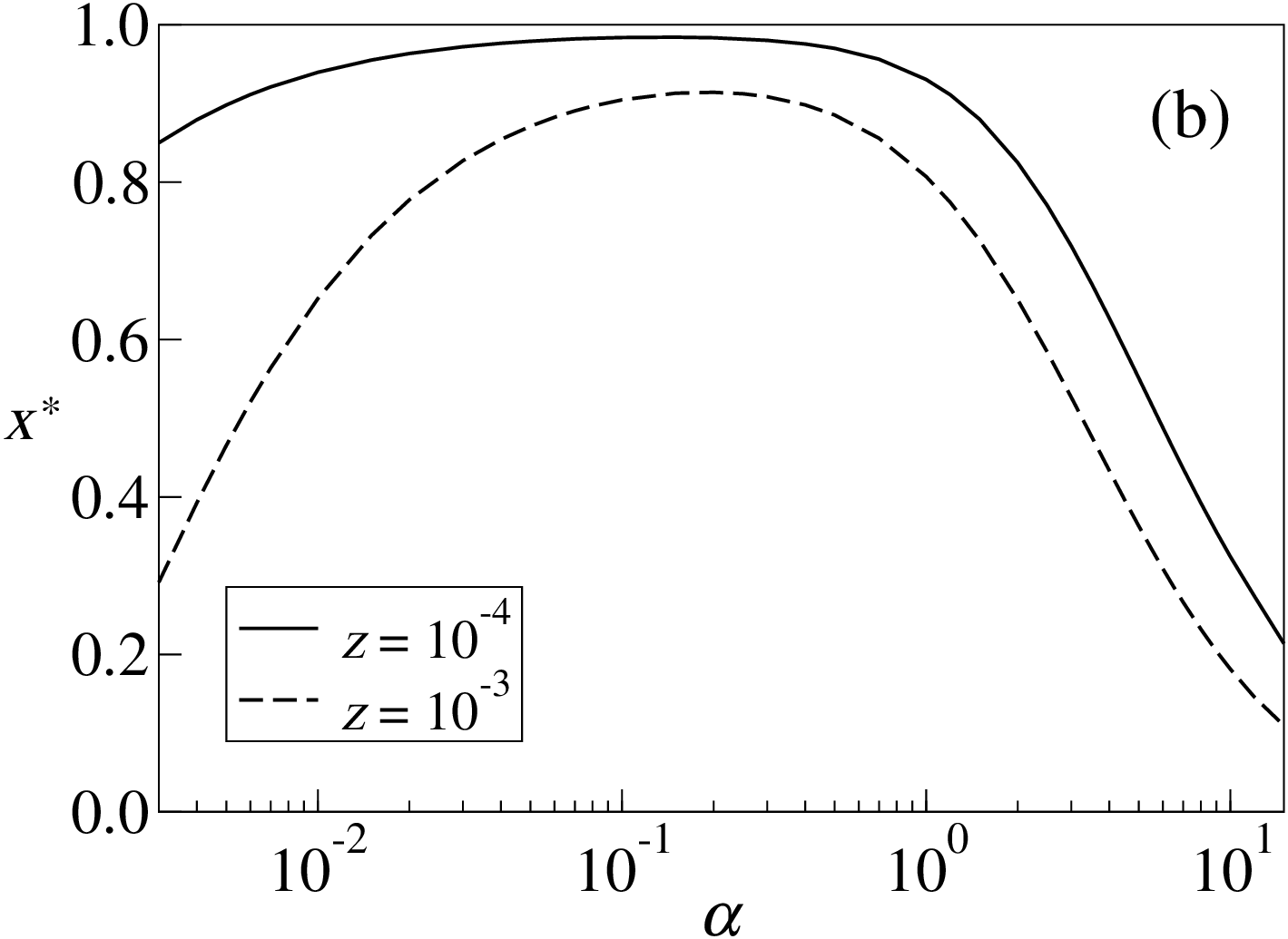}
\caption{Optimal quality (a) and optimal vendor profit per buyer
(b) versus acceptance parameter $\alpha$.}
\label{fig:spam}
\end{figure}

By the economics of spamming we understand the situation when
a low initial cost $Z$ allows the vendor to produce an abundance
of low quality variants. We simplify our considerations to $M$
variants with identical qualities $Q$ and identical buyers with
acceptance parameter $\alpha$. Assuming that the variants
are displayed consecutively, the selection probability plays no
role. This situation resembles spam messages arriving into our
mailboxes which we reject one by one. According to
Eq.~\req{Paccept}, on average $N(1-[1-P_A(Q,\alpha)]^M)$ buyers
accept one of the displayed variants and the expected
vendor profit per buyer is therefore
\begin{equation}
\label{X-spam}
x(Q,M)=(1-Q)\Big(1-\big[1-P_A(Q,\alpha)\big]^M\Big)-Mz.
\end{equation}
Since we focus on low quality variants, $P_A(Q,\alpha)$ is small
and the approximation $1-x\approx\exp[-x]$ can be used. It
follows that for a given quality $Q$, the optimal number of
displayed variants is
\begin{equation}
\label{Mopt}
M^*(Q)=\frac{\alpha+1}{\alpha Q^{\alpha}}
\ln\frac{\alpha(1-Q)Q^{\alpha}}{z(\alpha+1)}.
\end{equation}
However, when buyers' perception is limited to a certain number
of variants $M_m$, this value applies instead of $M^*(Q)$.
Assuming small initial cost $z$, the leading contribution to the
optimal quality can be shown to verify
$b\ln[\alpha Q^{\alpha}/b]=Q^{\alpha+1}$, where
$b:=z(1+1/\alpha)$). When the resulting $Q^*\ll1$, it follows
that Eq.~\req{Mopt} take the simple form
$M^*(Q^*)=Q^*/(\alpha z)$; the optimal quality
$Q^*$ has to be found numerically. The results are shown in
Fig.~\ref{fig:spam}: as $z$ gets lower, the optimal quality
decreases and the optimal profit increases; in addition, when
acceptance parameters are small
($10^{-2}\lesssim\alpha\lesssim1$) the optimal profit per buyer
is almost one which means that nearly all buyers react to the
spamming.


\begin{thebibliography}{99}
\bibitem{Ak70} G. Akerlof,
\emph{Quarterly Journal of Economics} \textbf{84}, 1970
(488--500)

\bibitem{Ho52} H. S. Houthakker,
\emph{The Review of Economic Studies} \textbf{19}, 1952
(155--164)

\bibitem{Sha82} C. Shapiro,
\emph{The Bell Journal of Economics} \textbf{13}, 1982 (20--35)

\bibitem{Zha01} Y.-C. Zhang,
\emph{Physica A} \textbf{299}, 2001 (104--120)

\bibitem{Mot93} M. Motta,
\emph{The Journal of Industrial Economics} \textbf{41}, 1993
(113--131)

\bibitem{Joh03} J. P. Johnson, D. P. Myatt,
\emph{The American Economic Review} \textbf{93}, 2003 (748--774)

\bibitem{ChaRo89} P. Champsaur, J.-Ch. Rochet,
\emph{Econometrica} \textbf{57}, 1989 (533--557)

\bibitem{BeWa03} S. Berry, J. Waldfogel,
\emph{NBER Working Paper No. 9675}, 2003

\bibitem{Sut91} J. Sutton,
\emph{Sunk Costs and Market Structure}, MIT Press, 1991

\bibitem{CaPe05} D. W. Carlton, J. M. Perloff,
\emph{Modern Industrial Organization (4rd edition)},
Addison-Wesley, 2005

\bibitem{McF80} D. McFadden,
\emph{Journal of Business} \textbf{53}, 1980 (13--29)

\bibitem{Cu82} I. S. Currim,
\emph{Journal of Marketing Research} \textbf{19}, 1982 (208--222)

\bibitem{AP92} S. P. Anderson, A. de Palma,
\emph{Oxford Economic Papers} \textbf{44}, 1992 (51--67)

\bibitem{MZ07} M. Medo and Y.-C. Zhang,
\emph{Physica A} \textbf{387}, 2008 (2889--2908)

\bibitem{Schm86} R. Schmalensee,
in \emph{New Developments in the Analysis of Market
Structure} (eds. J. E. Stiglitz and G. F. Mathewson),
MIT Press, 1986

\bibitem{Zha05} Y.-C. Zhang,
\emph{Physica A} \textbf{350}, 2005 (500--532)

\bibitem{McF74} D. McFadden,
\emph{Frontiers in Econometrics} \textbf{8}, 1974 (105--142)

\bibitem{McA07} R. P. McAfee,
\emph{Economics: The Open-Access, Open-Assessment E-Journal}
\textbf{1}, 2007-1

\bibitem{MuRo78} M. Mussa, S. Rosen,
\emph{Journal of Economic Theory} \textbf{18}, 1978 (301--317)
\end{thebibliography}
\end{document}